\documentclass[twocolumn]{aastex631}

\usepackage{multirow}
\usepackage{graphicx}
\usepackage{tablefootnote}
\usepackage{footnote}
\usepackage{txfonts}
\usepackage{tabularx}
\usepackage{appendix}
\usepackage{threeparttable}  
\usepackage{dcolumn}

\usepackage{amsmath}
\usepackage{longtable}
\usepackage{float}
\usepackage{hyperref}
\usepackage{footmisc}
\usepackage{booktabs} 

\def\cwt#1 {{\textcolor{purple}{#1}}\ }

\received{: 3-Jun-2024}
\revised{:  27-Aug-2024}
\accepted{: 17-Sep-2024}

\begin{document}

\title{The Variability of Persistent Radio Sources of Fast Radio Bursts}

\correspondingauthor{Feng Yi, Di Li, Chao-Wei Tsai}
\email{yifeng@zhejianglab.com, dili@nao.cas.cn, cwtsai@nao.cas.cn}

\author[0000-0003-4546-2623]{Ai Yuan Yang}
\affil{National Astronomical Observatories, Chinese Academy of Sciences, Beijing 100101, China}
\affiliation{Key Laboratory of Radio Astronomy and Technology, Chinese Academy of Sciences, A20 Datun Road, Chaoyang District, Beijing, 100101, P. R. China}

\author[0000-0002-0475-7479]{Yi Feng}
\affil{Research Center for Astronomical Computing, Zhejiang Laboratory, Hangzhou 311100, China}

\author[0000-0002-9390-9672]{Chao-Wei Tsai}
\affil{National Astronomical Observatories, Chinese Academy of Sciences, Beijing 100101, China}
\affiliation{Institute for Frontiers in Astronomy and Astrophysics, Beijing Normal University, Beijing 102206, China}
\affiliation{Key Laboratory of Radio Astronomy and Technology, Chinese Academy of Sciences, A20 Datun Road, Chaoyang District, Beijing, 100101, P. R. China}

\author[0000-0003-3010-7661]{Di Li}
\affil{Department of Astronomy, Tsinghua Univerisity, 30 Shuangqing Road, Beijing 100084, China}
\affil{National Astronomical Observatories, Chinese Academy of Sciences, Beijing 100101, China}
\affiliation{Key Laboratory of Radio Astronomy and Technology, Chinese Academy of Sciences, A20 Datun Road, Chaoyang District, Beijing, 100101, P. R. China}

\author{Hui Shi}
\affil{National Astronomical Observatories, Chinese Academy of Sciences, Beijing 100101, China}
\affiliation{Key Laboratory of Radio Astronomy and Technology, Chinese Academy of Sciences, A20 Datun Road, Chaoyang District, Beijing, 100101, P. R. China}

\author[0000-0002-3386-7159]{Pei Wang}
\affil{National Astronomical Observatories, Chinese Academy of Sciences, Beijing 100101, China}
\affiliation{Institute for Frontiers in Astronomy and Astrophysics, Beijing Normal University, Beijing 102206, China}
\affiliation{Key Laboratory of Radio Astronomy and Technology, Chinese Academy of Sciences, A20 Datun Road, Chaoyang District, Beijing, 100101, P. R. China}
\author{Yuan-Pei Yang}
\affil{South-Western Institute for Astronomy Research, Yunnan University, Kunming 650504, China}
\affil{Purple Mountain Observatory, Chinese Academy of Sciences, Nanjing 210023, China}
\author{Yong-Kun Zhang}
\affil{National Astronomical Observatories, Chinese Academy of Sciences, Beijing 100101, China}
\author{Chen-Hui Niu}
\affil{Institute of Astrophysics, Central China Normal University, Wuhan 430079, Hubei, China}
\author{Ju-Mei Yao}
\affil{Xinjiang Astronomical Observatory, Chinese Academy of Sciences, Urumqi, Xinjiang 830011, China}
\author{Yu-Zhu Cui}
\affil{Research Center for Astronomical Computing, Zhejiang Laboratory, Hangzhou 311100, China}
\author{Ren-Zhi Su}
\affil{Research Center for Astronomical Computing, Zhejiang Laboratory, Hangzhou 311100, China}
\author{Xiao-Feng Li}
\affil{Changzhou Institute of Technology, 666 Liaohe Road, Changzhou 213002, People's Republic of China}
\author{Jun-Shuo Zhang}
\affil{National Astronomical Observatories, Chinese Academy of Sciences, Beijing 100101, China}
\author{Yu-Hao Zhu}
\affil{National Astronomical Observatories, Chinese Academy of Sciences, Beijing 100101, China}
\affil{University of Chinese Academy of Sciences, Beijing 100049, People's Republic of China}
\author{W. D. Cotton}
\affil{National Radio Astronomy Observatory, 520 Edgemont Road, Charlottesville, VA 22903, USA}

\begin{abstract}
Over 700 bright millisecond-duration radio transients, known as Fast Radio Bursts (FRBs), have been identified to date. Nevertheless, the origin of FRBs remains unknown. 
The two repeating FRBs (FRB~20121102A and FRB~20190520B) have been verified to be associated with persistent radio sources (PRSs), making them the best candidates to study the nature of FRBs. 
Monitoring the variability in PRSs is essential for understanding their physical nature.
We conducted 22 observations of the PRSs linked to FRB~20121102A and FRB~20190520B using the Karl G. Jansky Very Large Array (VLA), to study their variability. We have observed significant flux variability for the PRSs of FRB~20121102A and FRB~20190520B, with a confidence level exceeding 99.99\%, based on the observations covering the longest timescale recorded to date. 
The observed variability of the two PRSs exhibits no significant difference in amplitude across both short and long timescales.
We found that the radio-derived star formation rates of the two FRB hosts are significantly higher than those measured by the optical $H_{\alpha}$ emissions, indicating that their host galaxies are highly obscured or most radio emissions are not from star formation processes. 
The observed timescale of PRS flux evolution constrained the magnetic field of FRB~20121102A with $B_\parallel\gtrsim1~{\rm mG}$ and FRB~20190520B with $B_\parallel\gtrsim0.1~{\rm mG}$.

\end{abstract}


\section{Introduction} \label{sec:intro}
Fast radio bursts (FRBs) are bright millisecond-duration radio transients, first discovered by \citet{Lorimer2007Sci318777L}. 
Since then, more than 700 FRBs have been reported to date \citep[e.g.,][]{CHIME_FRB2021ApJSCat,TheCHIME_FRB_cat2023arXiv231100111T}, as summarized by \citet{Xu2023Univ9330X} for the latest FRB database named Blinkverse\footnote{\url{https://blinkverse.alkaidos.cn}}. However, their progenitors and radiation mechanisms remain unknown. 
Most of the discovered FRBs are observed as non-repeating events, referred to as one-time events or on-offs, which makes it challenging to study the nature of FRBs in detail. 
The discovery of repeating FRBs has significantly influenced our understanding of the origin of FRBs because the repetition makes the sources easier to localize \citep[e.g.,][]{Law2022ApJ92755L}. 
To date, there are three repeat FRBs are found to be associated with persistent radio sources that have been precisely located \citep{Chatterjee2017Natur54158C,Niu2022Natur606873N,Li2021Natur267L,Feng2022Sci3751266F,Bruni2023arXiv231215296B}. 
The first repeating FRB, known as FRB~20121102A, was also the first found to be associated with a luminous persistent radio source (PRS) that has been localized to have arcsecond precision \citep{Chatterjee2017Natur54158C}, with a distance of $d$ = 972\,Mpc and a redshift $z= 0.193$ \citep{Tendulkar2017ApJ834L7T}. FRB~20121102A is likely located in the complex environment with circular polarization detected \citep{121102rm, feng2022b}. 
The second repeating FRB~20190520B was discovered by \citet{Niu2022Natur606873N} with the Five-hundred-meter Aperture Spherical radio Telescope \citep[FAST;][]{Nan2011IJMPD20989N,Li2019RAA1916L}, which is found to have a distance $d$ = 1218\,Mpc, a dispersion measure $\rm DM=1205 \pm 4\,pc\,cm^{-3}$, and a redshift $z= 0.241$ for its dwarf host galaxy (J160204.31$-$111718.5). FRB~20190520B is also likely located in a complex environment \citep{feng2023, feng2022a}. 
Recently, \citet{Bruni2023arXiv231215296B} reported the third FRB associated with the nearby FRB~20201124A at a distance of 413\,Mpc, showing a low luminosity with nebular origin.   
These persistent radio sources with precision location are the best targets to studying the host of FRBs. 

 The PRSs of these repeating FRBs could be explained by the non-thermal radiation from a nebula, e.g., a supernova remnant or a pulsar wind nebulae, surrounding a magnetar \citep[e.g.,][]{Murase2017ApJ836L6M,Yang2016ApJ819L12Y,Beloborodov2017ApJ843L26B,Metzger2017ApJ84114M,Margalit2018ApJ868L4M} 
or an accreting compact object 
\citep[e.g., ][]{Michilli2018Natur553182M,Zhang2018ApJ854L21Z,Sridhar2022ApJ9375S,Sridhar2024ApJ96074S}. 
For example, \citet{Yang2016ApJ819L12Y} proposed that the PRS could be produced by the synchrotron-heating process from radio bursts in a self-absorbed synchrotron nebula, and the observed spectral feature of the PRS associated with FRB~20121102A could be well explained by such a scenario \citep{Li2020ApJ89671L}. \citet{Dai2017ApJ838L7D} suggested that the PRS could be generated via a relativistic pulsar wind nebula sweeping up its ambient medium. On the other hand, it is noteworthy that all three repeating FRBs with observed PRSs have large Faraday rotation measures (RMs) \citep[e.g.,][]{Michilli2018Natur553182M,Niu2022Natur606873N,AnnaThomas2023Sci380599A}
which implies that observed RM mostly arises from the PRS region. If relativistic electrons constitute a significant fraction of the medium in the RM region, the magnetized environment surrounding the FRB source would produce synchrotron radiation, powering a bright PRS, meanwhile, there would be a simple relation between the PRS luminosity and the RM \citep[e.g.,][]{Yang2020ApJ8957Y,Yang2022ApJ928L16Y,Bruni2023arXiv231215296B}.

These current models face challenges in response to the growing diversity in the observed properties of FRBs and their host environments. 
For instance, the models connected FRB hosts to `magnetars', which are neutron stars with extremely powerful magnetic fields \citep[e.g.,][]{Margalit2019ApJ886110M}, are supported by the detection of an FRB-like event (FRB~20200428A) from the Galactic magnetar SGR\,1935+2154 \citep{Bochenek2020Natur58759B,CHIME_FRB_Collaboration2021ApJS25759C}. 
Addtionally, a small subset of the FRB hosts is found to be associated with quiescent and massive galaxies \citep{Li2020ApJ899L6L,Gordon2023ApJ95480G,Sharma2023ApJ950175S}, while the majority hosts are star-forming galaxies \citep{Bhandari2023ApJ94867B,Gordon2023ApJ95480G}, including the two repeating FRBs of this work hosted by star-froming dwarf galaxies  \citep{Chatterjee2017Natur54158C,Niu2022Natur606873N}. 
The above information suggests that the FRB sources
may originate from various progenitor systems with multiple-origin hypotheses \citep[e.g., ][]{Law2022ApJ92755L}. 
To explore the potential origins of FRBs with hosts in star-forming galaxies, the repeating FRBs associated with PRSs investigated in this study are the best cases.

The study of variability in radio sources is thought to be connected to the nature of their host galaxies \citep[e.g.][]{Cotton1976ApJS32467C,Cotton1983ApJ27151C,Bell2014MNRAS438352B,Sarbadhicary2021ApJ92331S}. 
The intensity variabilities of the PRSs associated with FRB~20121102A and FRB~20190520B have been discussed \citep{Chatterjee2017Natur54158C,Plavin2022MNRAS5116033P,Zhang2023ApJ95989Z}, however, the properties of their flux variability in the references are far from being concluded. For instance,  
\citet{Chatterjee2017Natur54158C} found a flux density variability of 10\% at 3\,GHz on the daily timescales for the PRS associated with FRB~20121101A, however, \citet{Plavin2022MNRAS5116033P} suggested that the variability at 1.7\,GHz and 4.8\,GHz is not significant and below 10\%. For the PRS associated with FRB~20190520B, \citet{Zhang2023ApJ95989Z} detected overall marginal variability with 3.2$\sigma$ radio flux decrease between the observations taken in 2020 and 2021 at 3\,GHz, while no significant variations were found at 1.5, 5.5, and 10\,GHz. 
Statistically, it has been observed that the uncertainties of the significance and degree of variability of a source exhibit a negative correlation with the number of observation epochs  \citep{Sarbadhicary2021ApJ92331S}. 
Therefore, a large number of observations with long-term intensity monitoring are required to achieve a statistically robust analysis of the variability for the two PRSs.

In this work, we conducted a series of VLA observations toward the PRSs of FRB~20121102A and FRB~20190520B to investigate their radio physical properties and monitor their fluxes to statistically discuss the significance and degree of the variability.  
In Section\,\ref{sect:obs_data}, we describe the observations, data calibration, and imaging of this study. Section\,\ref{sect:results} presents the observational results and the investigation of the variability in the radio light curve using several methods. 
The environment of the FRB hosts is discussed in Sect\,\ref{sect:discussion}. We draw our conclusion in Section\,\ref{sect:conclusion}. 

\section{Observations and data reduction}
\label{sect:obs_data}
\subsection{Observations}
\label{sec:obs}
The continuum observations were carried out using VLA (project ID: 23A-010, PI: Feng Yi), covering L-band 1-2\,GHz and the S-band 2-4\,GHz between May and July of 2023. 
In total, 10 and 12 observation scheduling blocks have been conducted for the L-band and S-band, respectively, with 16 spectral windows and 64 channels, each channel having a bandwidth of 2\,MHz.
The synthesized beam in B, BnA, and BnA$\to$A configuration at L-band (1--2\,GHz) and S-band (2--4\,GHz) is $\sim4.0\arcsec$ and $\sim 2\arcsec$, respectively. 
The phase calibrators were observed every 10 minutes to correct the amplitude and phase of the interferometer data for atmospheric and instrumental effects.
The absolute flux density scale was calibrated by comparing the standard flux calibrators 3C\,286 and 3C\,147 with their models provided by the NRAO \citep{Perley2017ApJS2307P}. 
The observation and instrument parameters in this work are summarized in Table\,\ref{tab:obs_info}.

\setlength{\tabcolsep}{2pt}
\begin{table}
\centering 
\caption{Summary of continuum observations using VLA and basic information for the PRSs of FRB~20121102A and FRB~20190520B.}
\label{tab_obsparms}
\begin{tabular}{l|c|c}\hline \hline
 
\multirow{2}{*}{Parameter} &   PRS of  &  PRS of     \\
&  FRB~20121102A & FRB~20190520B  \\
\hline
\multicolumn{3}{c}{VLA continuum observation} \\
\hline
Project              & \multicolumn{2}{c}{23A-010} \\
\hline
Frequency (GHz) & L-band (1-2\,GHz) &  S-band (2-4\,GHz)  \\
\hline
Array configuration  & \multicolumn{2}{c}{B, BnA, BnA$\to$A} \\  
\hline
Observing mode       & \multicolumn{2}{c}{Continuum} \\ 
\hline
Spectral window      &  \multicolumn{2}{c}{16} \\
\hline
No. Channels         &  \multicolumn{2}{c}{64} \\ 
\hline
Bandwidth per channel      & \multicolumn{2}{c}{2\,MHz} \\ 
\hline
Primary beam         & $\sim 30 \arcmin$ & $15\arcmin$ \\ 
\hline
Synthesized beam     & $\sim 4\arcsec$ & $\sim 2 \arcsec$  \\
\hline
Observing dates      & \multicolumn{2}{c}{2023 May. $-$ 2023 Jul.} \\ 
\hline
Observation schedule blocks   & 10 & 12 \\
\hline
Total time per observation & \multicolumn{2}{c}{$\sim$ 1.0\,hrs} \\ 
\hline
Typical sensitivity      & $\rm \sim 50\,\mu Jy$ & $\rm \sim 15\,\mu Jy$  \\
\hline
Total observing time & $\sim$10\,h & $\sim$12\,h \\  
\hline
Flux density calibrator (Jy) &  3C\,147 & 3C\,286  \\ 
\hline 
\multicolumn{3}{c}{Basic information} \\
\hline
Distance (Mpc) & 972 &  1218 \\
\hline
Redshift $z$ & 0.193  & 0.241  \\
\hline
\hline
\end{tabular}
\label{tab:obs_info}
\begin{tablenotes}
\item {Note: The BnA configuration is a hybrid of the B and A configurations (i.e., combining 1.0 $\times$ [larger configuration] + 0.4 $\times$ [smaller configuration].), while Bn$\to$A refers to data taken during the transition from the BnA to A configuration. } 
\end{tablenotes}
\end{table}

\begin{figure*}[!hpt]
 \centering
   \begin{tabular}{cc}
   \includegraphics[width = 0.5\textwidth]{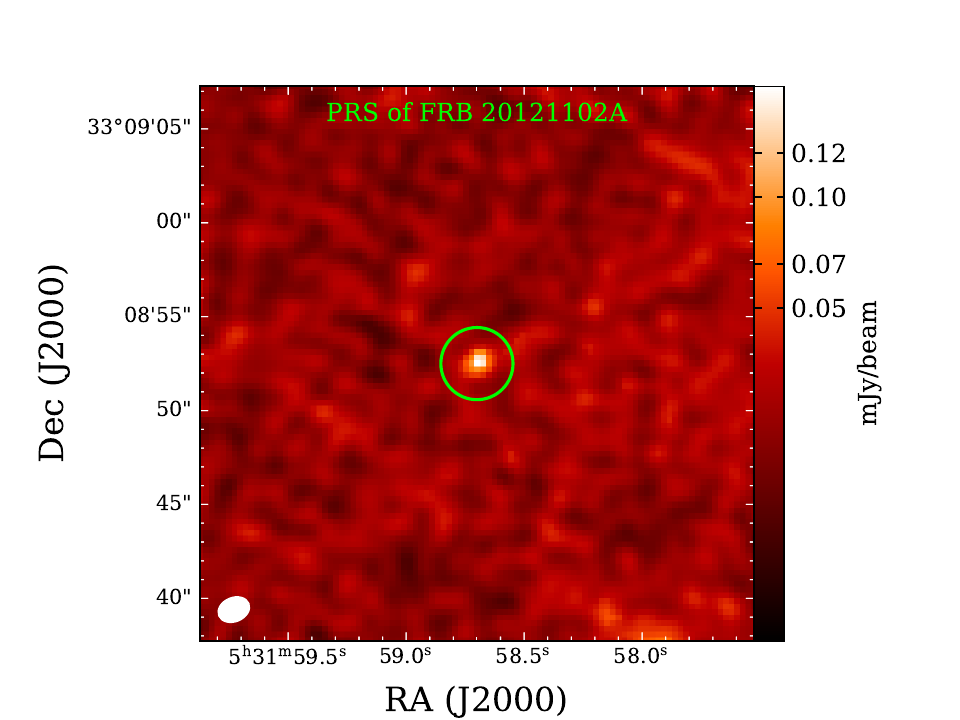}
 & 
 \hspace{-5mm}\includegraphics[width = 0.5\textwidth]{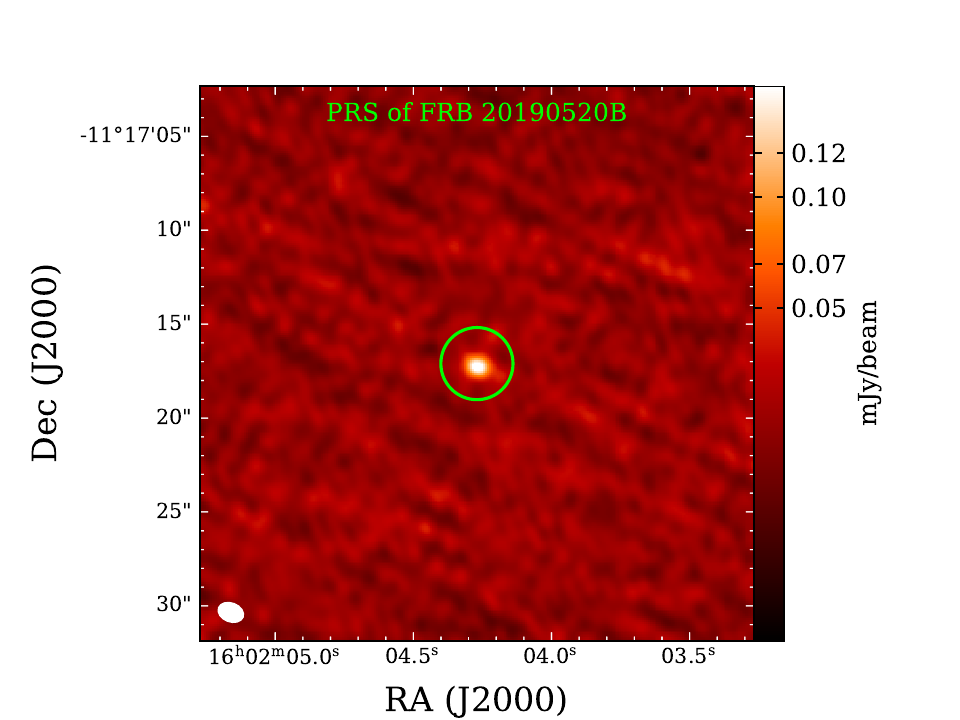}\\
    \end{tabular}
    \caption{ Left panel: The L-band continuum image for the PRS of FRB~20121102A. 
     Right panel: The S-band continuum image for the PSR of FRB~20190520B. The FWHM beam is indicated by the white circles in the lower-left corner of each image. The PRS of each FRB are compact and unresolved compared to the beam size. }
 \label{fig:FRB_img}%
\end{figure*}
\subsection{Data reduction} 
The data reduction pipelines were developed from the continuum pipelines used in B-configuration of the GLOSTAR survey \citep[e.g., ][]{Brunthaler2021AA651A85B}, which made use of the OBIT\,\footnote{\url{https://www.cv.nrao.edu/~bcotton/Obit.html}} package \citep{Cotton2008PASP439C} with scripts written in \emph{python} to access tasks from the Astronomical Image Processing Software package  \citep[AIPS\footnote{\url{https://www.aips.nrao.edu/}},][]{Greisen2003ASSL109G}. 
The full details of data calibration and imaging can be found in the B-configuration catalog papers of the GLOSTAR survey \citep[e.g., ][]{Dzib2023AA670A9D,Yang2023arXiv231009777Y}. 

In brief, the raw data was calibrated and edited following 12 main steps, to flag bad data and correct amplitude and phase in time and frequency. 
Each calibration step was followed by an editing step looking for deviant solutions and flagging the corresponding data.
These corrected flags were kept and automatically added to the flag lists. 
The additional flags from the checking of diagnostic plots of various stages were also manually added to the flag lists. 
The flag lists were applied when the calibration pipeline was redone. 
Since the calibration pipeline was carried out iterative, most of the radio frequency interferences (RFIs) were removed by the pipeline. 
Combined with the manual checks, the effects of RFIs in the L-band and S-band observations of this work were minimized. 
The calibrated datasets were then used to be cleaned in the imaging process, which first did a shallow CLEAN to get a crude sky model and then a deeper multifrequency CLEAN.  
During the imaging process, each field of view consists of multiple facets that are cleaned in parallel. These cleaned facets are then combined into a single plane for each subband.

\begin{figure*}[!hpt]
 \centering
   \includegraphics[width = 0.85\textwidth]{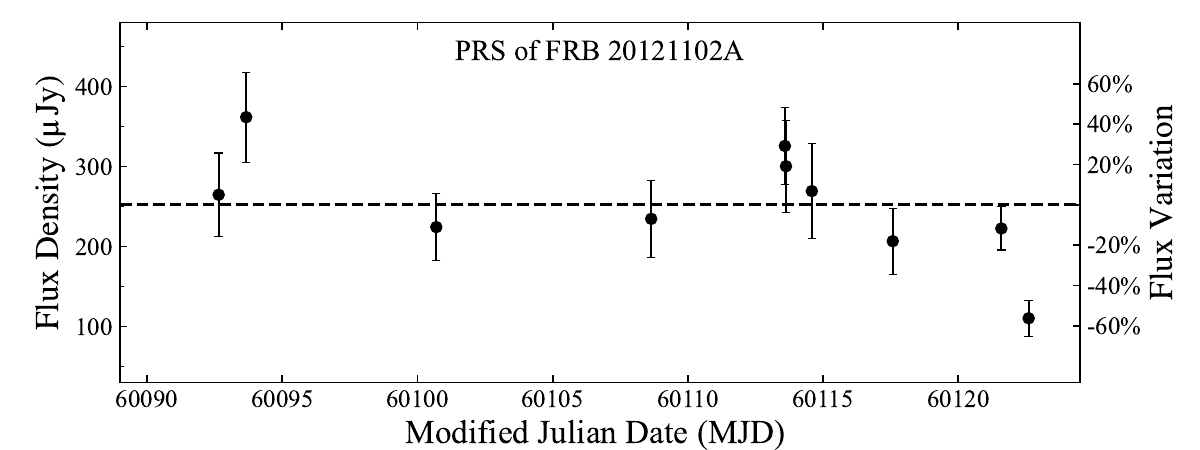} \\
   \includegraphics[width = 0.85\textwidth]{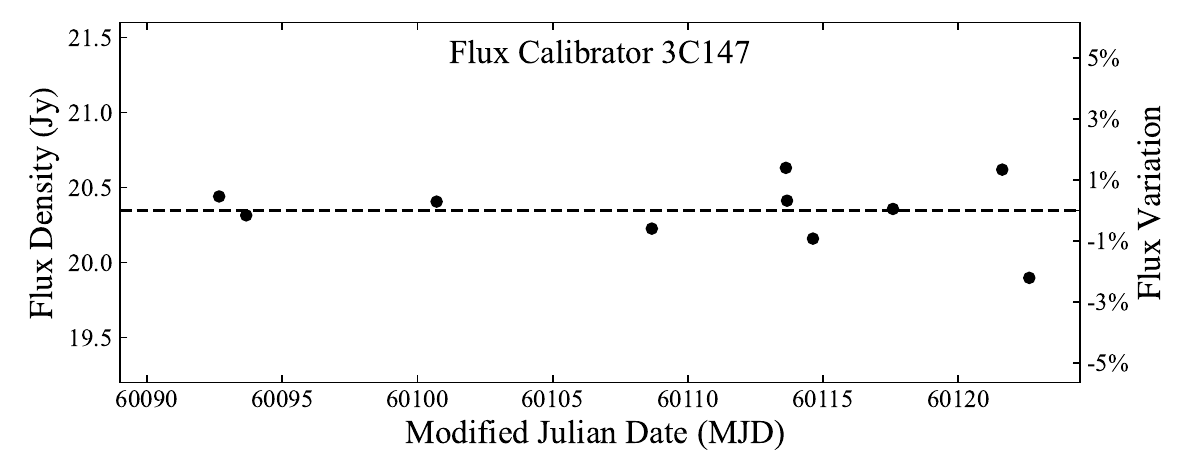}
    \caption{The light curves of the PRS associated with FRB~20121102A (top panel) and the flux calibrator 3C\,147 (bottom panel) at the L-band 1.5\,GHz.  The dashed line shows the mean value of flux density for each source. The flux variation in the flux calibrator 3C\,147 is less than three percent (3\%), which is expected as discussed in \citet{Perley2017ApJS2307P}. The flux density variations of the PRS range from 12\% to 60\%, compared to the mean value in the light curve. No significant correlation was found for the flux density variation between the target and calibrator under Pearson’s correlation test with a $p$-value$>$0.1. Therefore, the flux changes in the PRS of FRB are reliable. 
     }
 \label{fig:light_curve_prs121102_fluxcal}%
\end{figure*}

\section{Results and Analysis} 
\label{sect:results}
The results from all 22 VLA observations present compact and unresolved persistent radio emission associated with FRB~20121102A at L-band 1.5\,GHz and FRB~20190520B at S-band 3\,GHz. These observations span from the end of May to the end of June 2023. The observational results of the 22 VLA detections of the work are listed in Table\,\ref{tab:obs_results_thiswork}.

\setlength{\tabcolsep}{8pt}
\begin{table*}
\centering
\caption{Observational results of PRSs of FRB\,20121102A and FRB\,20190520B in this work}
\begin{tabular}{lccccccl}
\hline
\hline
Obs. Date\,(Start Time) & Time  & Frequency &  RMS & Flux Density & Beam Size & Beam Position  &  Flux Calibrator \\
YY-MM-DD-HH & MJD & GHz & $\rm \mu Jy/beam$  & $\rm \mu Jy$  & $\arcsec \times \arcsec$ &  Angle ($\degr$) & Jy \\
\hline
(1) & (2)  & (3) & (4) & (5) & (6) & (7) & (8) \\
\hline
\multicolumn{8}{c}{PRS associated with  FRB~20121102A at L-band}     \\
\hline
2023-05-28T15:33:50 & 60092.6 & 1.5 & 48.7 & 264.7$\pm$52.2 & 3.8$\times$3.4 & -71.6 & 20.4 \\
2023-05-29T16:00:03 & 60093.7 & 1.5 & 60.2 & 361.6$\pm$56.4 & 3.4$\times$3.1 & -85.2 & 20.3 \\
2023-06-05T16:44:11 & 60100.7 & 1.5 & 44.8 & 224.3$\pm$41.6 & 2.9$\times$2.8 & 41.8 & 20.4 \\
2023-06-13T15:27:44 & 60108.6 & 1.5 & 51.0 & 234.6$\pm$47.8 & 4.0$\times$3.6 & 75.1 & 20.2 \\
2023-06-18T14:04:14 & 60113.6 & 1.5 & 53.5 & 325.7$\pm$48.0 & 4.5$\times$4.0 & 65.3 & 20.6 \\
2023-06-18T15:04:05 & 60113.6 & 1.5 & 57.6 & 300.4$\pm$57.3 & 4.6$\times$3.8 & 70.0 & 20.4 \\
2023-06-19T14:00:18 & 60114.6 & 1.5 & 62.6 & 269.2$\pm$59.5 & 4.8$\times$3.8 & 67.7 & 20.2 \\
2023-06-22T13:48:30 & 60117.6 & 1.5 & 41.0 & 206.6$\pm$40.5 & 4.4$\times$1.0 & -61.9 & 20.4 \\
2023-06-26T14:12:12 & 60121.6 & 1.5 & 29.1 & 222.5$\pm$28.7 & 2.0$\times$1.2 & -60.4 & 20.6 \\
2023-06-27T14:25:50 & 60122.6 & 1.5 & 23.9 & 110.3$\pm$22.7 & 2.1$\times$1.7 & -61.3 & 19.9 \\
\hline
\multicolumn{8}{c}{PRS associated with  FRB~20190520B at S-band}     \\
\hline
2023-06-07T07:14:53 & 60102.2 & 3.0 & 11.9 & 174.8$\pm$11.8 & 2.6$\times$1.2 & 57.5 & 10.0 \\
2023-06-15T05:06:31 & 60110.2 & 3.0 & 18.5 & 199.0$\pm$2.7 & 3.7$\times$2.9 & 1.3 & 9.9 \\
2023-06-17T04:06:46 & 60112.2 & 3.0 & 14.0 & 201.5$\pm$16.6 & 2.2$\times$1.5 & -5.5 & 10.0 \\
2023-06-18T03:45:56 & 60113.1 & 3.0 & 13.3 & 233.3$\pm$18.8 & 1.6$\times$1.6 & 19.0 & 9.9 \\
2023-06-18T04:45:46 & 60113.2 & 3.0 & 14.7 & 206.2$\pm$19.5 & 2.0$\times$1.9 & 41.8 & 10.1 \\
2023-06-20T03:38:04 & 60115.1 & 3.0 & 16.9 & 161.5$\pm$16.5 & 2.9$\times$1.9 & 11.4 & 9.7 \\
2023-06-20T04:37:54 & 60115.2 & 3.0 & 12.8 & 166.9$\pm$15.1 & 2.2$\times$1.3 & 23.1 & 9.7 \\
2023-06-20T05:37:45 & 60115.2 & 3.0 & 10.6 & 139.9$\pm$12.0 & 1.8$\times$1.2 & 21.4 & 10.0 \\
2023-06-20T06:37:41 & 60115.3 & 3.0 & 9.3 & 165.5$\pm$8.8 & 2.3$\times$1.7 & 24.9 & 9.8 \\
2023-06-23T05:47:50 & 60118.2 & 3.0 & 12.4 & 168.4$\pm$12.0 & 1.4$\times$1.0 & 69.0 & 9.7 \\
2023-06-24T03:31:28 & 60119.1 & 3.0 & 13.4 & 185.9$\pm$8.6 & 1.6$\times$1.3 & -67.2 & 10.0 \\
2023-06-24T04:31:20 & 60119.2 & 3.0 & 11.6 & 194.9$\pm$11.2 & 2.3$\times$0.9 & 80.4 & 9.9 \\
\hline
\hline
\end{tabular}
\label{tab:obs_results_thiswork}
\begin{tablenotes}
\item {Note: Column (1-2) lists the observation time in units of YY-MM-DD-HH and MJD. The center frequency, typical RMS noise, and the flux density of the target are shown in Columns 3-5. Columns 6-7 give the beam size and beam position angle. Please note that the differences in beam sizes are mainly due to the configurations of observations ranging from B, BnA, to BnA$\to$A, the elevation of the observation, and the first spectral window with enough good channels to do the image. Column 8 displays the flux density of the flux calibrator such as 3C\,147 at L-band and 3C\,286 at S-band.} 
\end{tablenotes}
\end{table*} 

\begin{figure*}[!hpt]
 \centering
   \begin{tabular}{cc}
   \includegraphics[width = 0.85\textwidth]{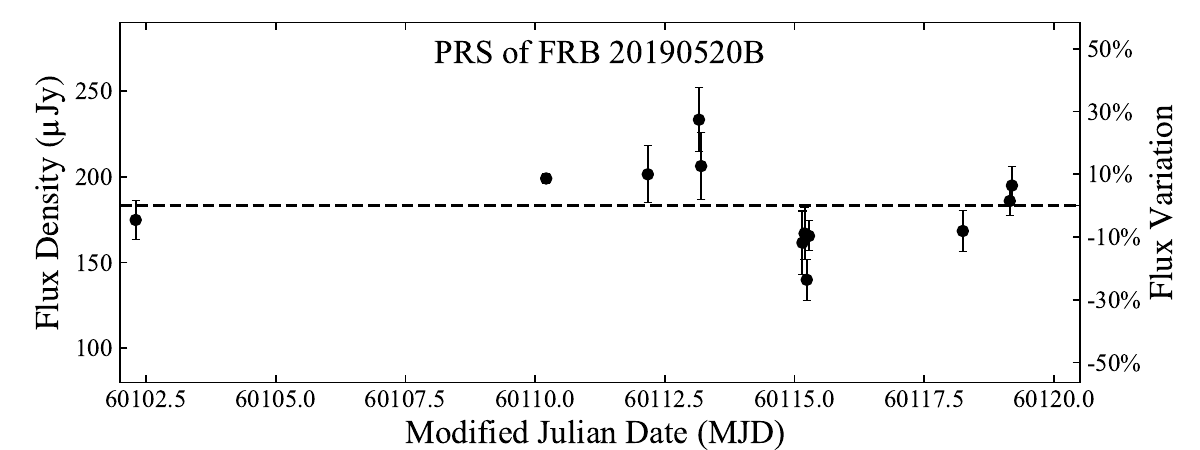} \\
   \includegraphics[width = 0.85\textwidth]{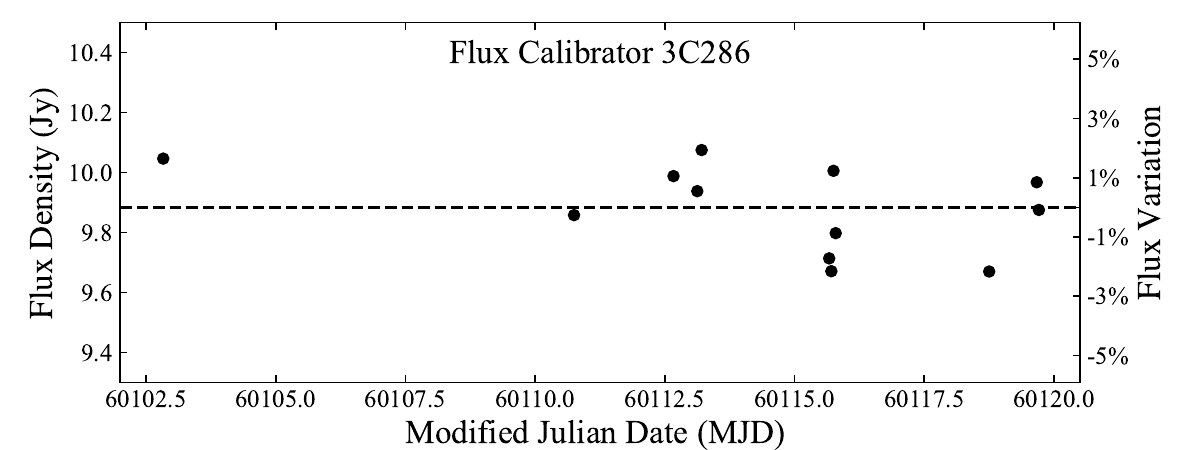} \\
    \end{tabular}
    \caption{ The light curves of the target FRB~20190520B (top panel) and the flux calibrator 3C\,286 (bottom panel) at the S-band 3\,GHz. {The dashed line shows the mean value of flux density for each source}. { The flux variation in the flux calibrator 3C\,286 is less than five percent (5\%)}, which is expected as discussed in \citet{Perley2017ApJS2307P}. 
    {The flux density variations of the PRS range from 3\% to 20\%, compared to the mean value in the light curve. }
    No significant correlation was found for the flux density variation between the target and calibrator under Pearson’s correlation test with a $p$-value$>$0.1.
    Therefore, the flux changes in the PRS of FRB are not related to the flux calibrator and are reliable. 
    }
 
 \label{fig:light_curve_prs190520_fluxcal}%
\end{figure*}

Fig.\,\ref{fig:light_curve_prs121102_fluxcal} and Fig.\,\ref{fig:light_curve_prs190520_fluxcal} show the flux densities of the flux calibrators and the persistent radio sources as a function of the observational dates in this work. 
As shown in the dashed lines for 5\% variation in the top panels of Fig.\,\ref{fig:light_curve_prs121102_fluxcal} and Fig.\,\ref{fig:light_curve_prs190520_fluxcal}, the flux densities variations in flux calibrators of 3C\,147 at the L-band and 3C\,286 at the S-band are less than 5\%, which is reasonable as discussed in \citet{Perley2017ApJS2307P} for VLA calibrators. 
Compared to the typical flux variation of $<5\%$ for the flux calibrators, much higher flux density variations ($20\%\sim 50\%$) are found in the light curve of two PRSs. Additionally, no significant correlations of the flux density variations are found between the targets and flux calibrators under Pearson’s correlation tests with $p$-values$>$0.1. 
Thus, the flux calibration of this study is affirmed to be reliable, and the observed variations in the targets are regarded as trustworthy. 

Table\,\ref{tab:obs_results_previous} lists the previous observational results for the persistent radio sources of FRB~20121102A and FRB~20190520B. For the observations with different observing frequencies of this work, we convert the flux densities to the frequency used in this work, based on the spectral index given in the references. 

\setlength{\tabcolsep}{12pt}
\begin{table*}[!htp]
\centering
\caption{Summary of the variability measurements of the two PRSs observed in this study.}
\begin{tabular}{lccccccccl}
\hline
\multicolumn{10}{c}{PRS of FRB\,20121102A} \\
\hline
& Num. of Obs. &  $\chi_{\rm lc}^{2}$   & $p(\chi_{\rm lc}^{2})$  & $m$ (\%) &   $m_{d}$ (\%)   & $V$(\%) &  $\eta$  & $V_{\rm flux}$ &  \\
\hline
& 27 &   125.4 & $<0.001$  & 21.7$\pm$1.8  & 21.5$\pm$1.8 & 22.1$\pm$1.8 & 82.6 & 0.39$\pm$0.05 & \\
\hline
\multicolumn{10}{c}{PRS of FRB\,20190520B} \\
\hline  
& Num. of Obs. &  $\chi_{\rm lc}^{2}$   & $p(\chi_{\rm lc}^{2})$  & $m$ (\%) &   $m_{d}$ (\%)   & $V$ (\%) & $\eta$  & $V_{\rm flux}$  &  \\
\hline
& 35 &    119.8 & $<0.001$ & 16.3 $\pm$1.7 & 16.0$\pm$1.8  & 16.4$\pm$1.7 & 207 & 0.21$\pm$0.05 & \\
\hline
\end{tabular}
\label{tab:var_summary}
\begin{tablenotes}[flush]
\item Note: To estimate the errors of the $m$, $m_{d}$, $V$, and $V_{\rm flux}$, we randomly generate flux in each observation by assuming that each measurement follows a Gaussian distribution with error of the measurement as the standard deviation of the distribution. The process was repeated 1000 times and the uncertainties of these parameters were estimated from the 68\% range (i.e., 1$\sigma$) of the distributions.
\end{tablenotes}
\end{table*}
\begin{figure*}[!hpt]
 \centering
   \includegraphics[width = 0.75\textwidth]{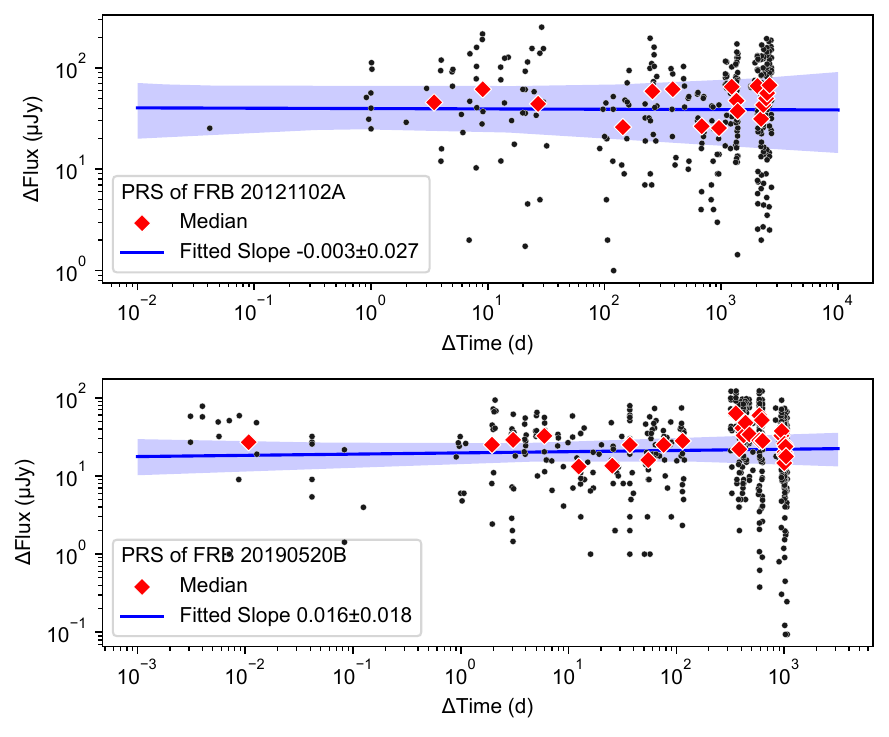}
    \caption{The flux differences ($\Delta$Flux) as a function of time differences ($\Delta$Time) for two different epochs of PRS of FRB~20121102A (top-panel) and FRB~20190520B (bottom-panel) are based on results in Table\,\ref{tab:obs_results_thiswork} and \ref{tab:obs_results_previous}.
    The solid blue line shows the linear fit for all points with $1\sigma$ shadows. The red diamonds show the median values derived from every 20 points in the figure. The figure displays both short and long timescales of variations for the PRSs of FRB~20121102A and FRB~20190520B.}
 \label{fig:diff_flux_vs_diff_time}%
\end{figure*}

\subsection{The PRS associated with FRB~20121102A} 
\label{subsec:frb121102}
Among the 10 VLA observations for the PRS of FRB~20121102A, we detected all of their emission at L-band. 
We conducted a 2D Gaussian fit for the cleaned images and obtained a point-like and unresolved source for the PRS, as outlined in \citealt{Yang2019MNRAS4822681Y,Yang2021AA645A110Y} for radio source. One example image of these observations is presented in the left panel of Fig.\,\ref{fig:FRB_img}.
The mean flux density is $\rm 251.9\pm66.6\rm\,\mu Jy$ for the 10 observations of the persistent source linked to FRB~20121102A at 1.5\,GHz, giving an average radio luminosity of $\rm L_{1.5\,GHz}\sim 4.2\times 10^{38}\,erg\,s^{-1}$. 
This is statistically consistent with the mean values of $\rm 228.2\pm40.8\,\mu Jy$ in the 13 previous observations (Table\,\ref{tab:obs_results_previous}), within the uncertainty.
From the 10 observations, we can see a clear variation in flux densities for the PRS, with minimum and maximum flux densities of $S_{\rm min}\pm\sigma_{\rm min}=110.3\pm22.7\,\mu$Jy to $S_{\rm max}\pm\sigma_{\rm max}=361.6\pm56.4\,\mu$Jy respectively, as shown in the Table\,\ref{tab:obs_results_thiswork}, suggesting a monthly flux variation.

\begin{figure}[!hpt]
 \centering
   \includegraphics[width = 0.5\textwidth]{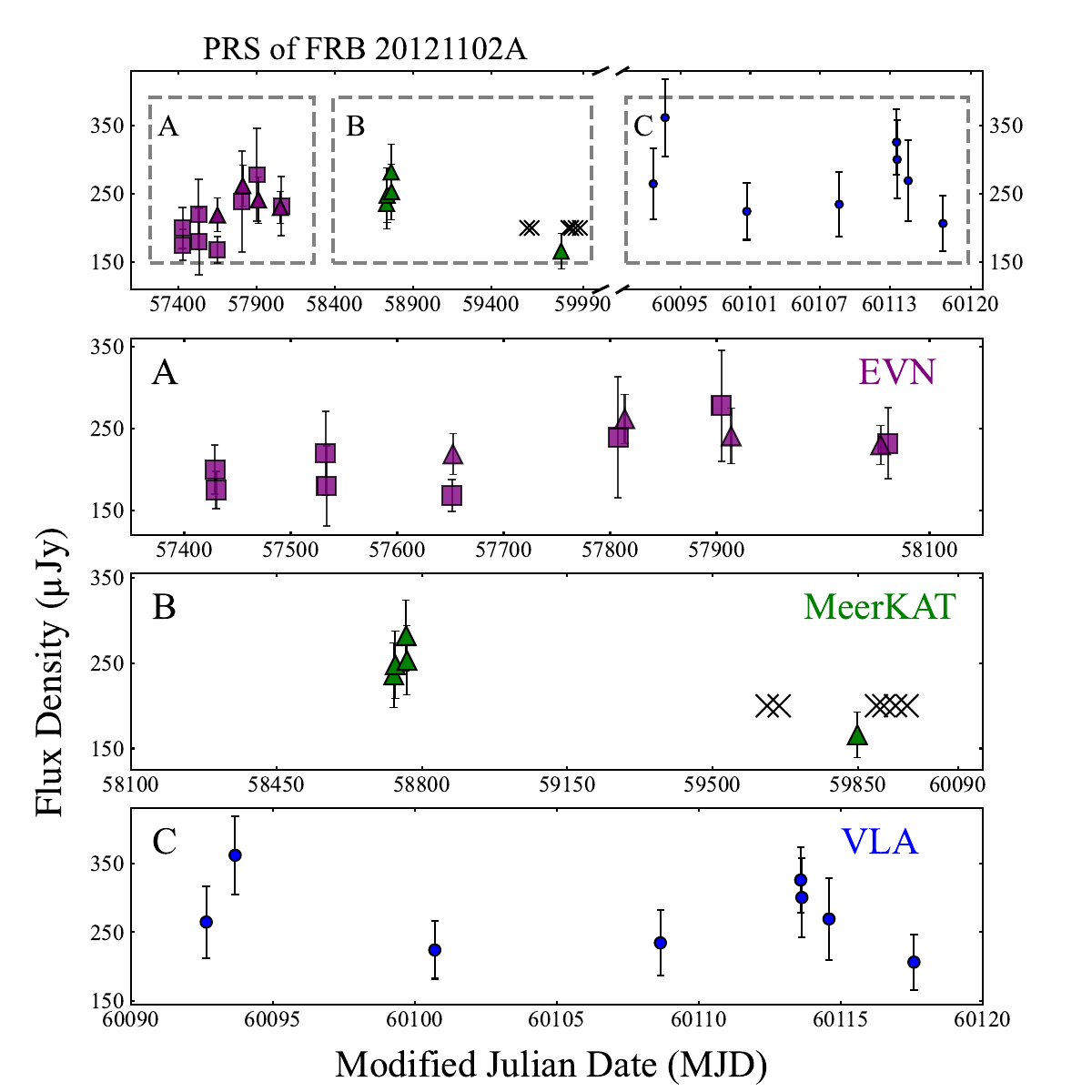} 
    \caption{Top-panel: The light curves of the PRS of FRB~20121102A at the L-band. The middle and bottom panels: A, B, and C show the zoom-in image of the corresponding regions in the top panel. 
    The black squares and triangles before MJD 60000 show the L-band flux densities measured/derived from the references \citep[e.g., ][]{Marcote2017ApJ834L8M,Plavin2022MNRAS5116033P,Rhodes2023MNRAS5253626R} which have summarized the previous observations \citep[e.g.,][]{Chatterjee2017Natur54158C}, and the dots represent the L-band flux densities in this work. The black crosses display the presence of observations that have not been made public.}
 \label{fig:light_curve_prs121102_t}%
\end{figure}

\subsection{The PRS associated with FRB~20190520B }
\label{subsec:frb190520}
The radio emission of the PRS linked to FRB~20190520B has been detected for the 12 VLA observations at the S-band. 
The 12 observations give a point-like or unresolved source for the PRS from the 2D Gaussian fit in the images, as shown in the right panel of Fig.\,\ref{fig:FRB_img}. 
The mean flux density is $\rm 183.2\pm24.1\rm\,\mu Jy$ for the 12 observations of the persistent source linked to FRB~20190520B at S-band, giving an average radio luminosity of $\rm L_{3.0\,GHz}\sim 2.7\times 10^{38}\,erg\,s^{-1}$.
The result is statistically consistent with the mean values of $\rm 175.8\pm30.0\rm\,\mu Jy$ in the 14 previous observations, within the uncertainty. 
The flux densities of the PRS are variable, ranging from $139.9\pm12.0\,\mu$Jy to $233.3\pm18.8\,\mu$Jy, indicating a monthly variation, as shown in the Table\,\ref{tab:obs_results_thiswork}. 


\begin{figure}[!hpt]
 \centering
   \includegraphics[width = 0.5\textwidth]{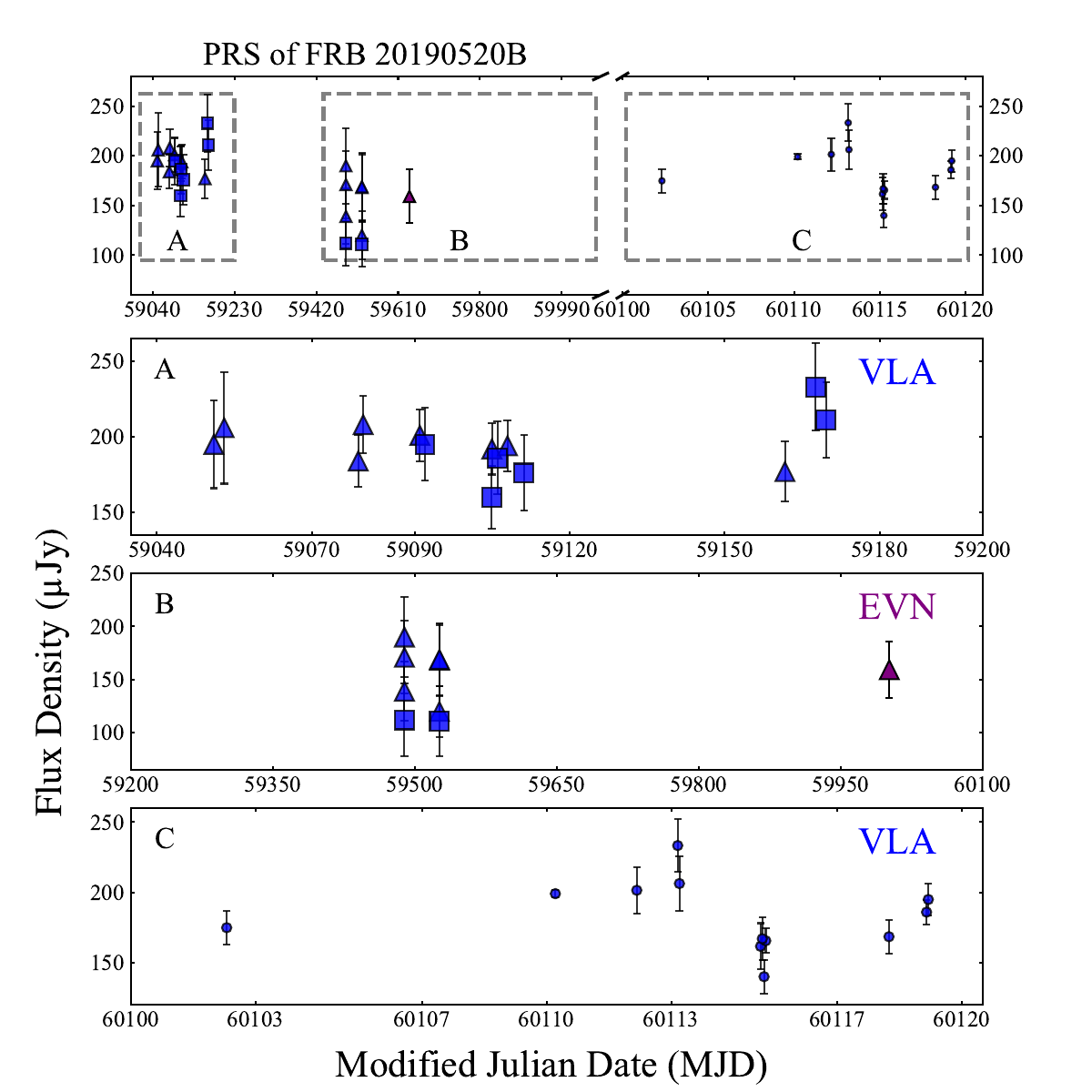} 
    \caption{ Top-panel: The light curves of the PRS of FRB~20190520B at 3\,GHz. The middle and bottom panels of A, B, and C show the zoom-in image of the corresponding regions in the top panel. The squares show the flux densities at 3\,GHz from the references \citep[e.g., ][]{Niu2022Natur606873N,Zhang2023ApJ95989Z,Bhandari2023ApJ958L19B}, and the dots represent the 3\,GHz flux densities in this work. The blue triangles display the measurements derived from other frequencies using the spectral index in the references \citep[e.g., ][]{Zhang2023ApJ95989Z}, as listed in Table\,\ref{tab:obs_results_previous}.}
 \label{fig:light_curve_prs190520_t}%
\end{figure}

\setlength{\tabcolsep}{2pt}
\begin{table*}
\centering
\caption{\large Previous observational results of PRSs associated with FRB~20121102A and FRB~20190520B.}
\begin{tabular}{lccc|cccc}
\hline
\hline 
\multicolumn{4}{c}{PRS associated with FRB~20190520B at S-band} &  \multicolumn{4}{|c}{PRS associated with  FRB~20121102A at L-band}   \\
\hline
Obs. Date  & Frequency  & Flux Density\,[spectral index] &  Ref. & Obs. Date & Frequency  & Flux Density\,[spectral index] &  Ref. \\
YY-MM-DD\,(MJD) & GHz & $\rm \mu Jy/beam$\,[$\alpha$] &   & YY-MM-DD\,(MJD)   & GHz &  $\rm \mu Jy/beam$\,[$\alpha$]  &  \\
\hline
2020-08-30\,(59092) &  3.0  & 195$\pm$24 &  (1)  & 2016-02-10\,(57428) & 1.7 & 200$\pm$20 &  (3)  \\
2020-09-12\,(59105) &  3.0  & 160$\pm$21 &  (1)  & 2016-02-11\,(57429) & 1.7 & 175$\pm$14 &  (3)  \\
2020-09-13\,(59106) &  3.0  & 186$\pm$24 &  (1)  & 2016-05-24\,(57532) & 1.7 & 220$\pm$40 &  (3) \\
2020-09-19\,(59111) &  3.0  & 176$\pm$25 &  (1)  & 2016-05-25\,(57533) & 1.7 & 180$\pm$40 &  (3) \\
2020-11-14\,(59167) &  3.0  & 233$\pm$29 &  (1)  &  2016-09-20\,(57651) & 1.7 & 168$\pm$11 &  (3) \\
2020-11-16\,(59169) &  3.0  & 211$\pm$25 &  (1)  & 2016-09-21\,(57652)$^{\ast}$ & 5.0$\to$1.5 & 123$\pm$14$\to$219[$\alpha=-0.48$] &  (3) \\
2021-10-01\,(59489) &  3.0  & 112$\pm$34 &  (2)  & 2017-02-23\,(57807) & 1.7 & 239$\pm$62 &  (4) \\
2021-11-07\,(59525) &  3.0  & 111$\pm$33 &  (2)  & 2017-03-01\,(57813)$^{\ast}$ & 4.8$\to$1.5 & 150$\pm$17$\to$262\,[$\alpha=-0.48$] &  (4) \\
2020-07-21\,(59051) & 1.5$\to$3.0  & 258$\pm$29$\to$195\,[$\alpha=-0.40$] &  (1)  & 2017-05-31\,(57904) & 1.7 & 278$\pm$54 &  (4) \\
2020-07-23\,(59053) & 1.5$\to$3.0  & 273$\pm$37$\to$206\,[$\alpha=-0.40$] &  (1)  & 2017-06-09\,(57913)$^{\ast}$ & 4.8$\to$1.5 & 138$\pm$22$\to$241\,[$\alpha=-0.48$] &  (4) \\
2021-10-01\,(59489) & 1.5$\to$3.0  & 240$\pm$70$\to$190\,[$\alpha=-0.33$] &  (2)  & 2017-10-27\,(58053)$^{\ast}$ & 4.8$\to$1.5 & 132$\pm$12$\to$230\,[$\alpha=-0.48$] &  (4) \\
2021-11-07\,(59525) & 1.5$\to$3.0  & 212$\pm$61$\to$168\,[$\alpha=-0.33$] &  (2)  & 2017-11-03\,(58060) & 1.7 & 232$\pm$32 &  (4) \\
2020-08-18\,(59079.0) & 5.5$\to$3.0  & 145$\pm$17$\to$184\,[$\alpha=-0.40$] &  (1)  & 2019-09-06\,(58732) & 1.28$\to$1.5 & 260$\pm$26$\to$236\,[$\alpha=-0.6$] &  (5) \\
2020-08-18\,(59079.9) & 5.5$\to$3.0  & 164$\pm$19$\to$208\,[$\alpha=-0.40$] &  (1)  & 2019-09-10\,(58736) & 1.28$\to$1.5 & 269$\pm$27$\to$248\,[$\alpha=-0.5$] &  (5) \\
2020-08-29\,(59090) & 5.5$\to$3.0  & 158$\pm$17$\to$201\,[$\alpha=-0.40$] &  (1)  & 2019-10-06\,(58762) & 1.28$\to$1.5 & 287$\pm$27$\to$282\,[$\alpha=-0.1$] &  (5) \\
2020-09-12\,(59105) & 5.5$\to$3.0  & 151$\pm$17$\to$192\,[$\alpha=-0.40$] &  (1)  & 2019-10-08\,(58764) & 1.28$\to$1.5 & 270$\pm$28$\to$253\,[$\alpha=-0.4$] &  (5) \\
2020-09-15\,(59108) & 5.5$\to$3.0  & 153$\pm$17$\to$194\,[$\alpha=-0.33$]  &  (1)  & 2022-09-26\,(59848) & 1.28$\to$1.5 & 189$\pm$18$\to$166\,[$\alpha=-0.8$] &  (5) \\
2020-11-08\,(59161) & 5.5$\to$3.0  & 139$\pm$20$\to$177\,[$\alpha=-0.33$] &  (1)  &  $-$  & $-$  & $-$   & $-$  \\
2021-10-01\,(59488) & 5.5$\to$3.0  & 114$\pm$28$\to$139\,[$\alpha=-0.33$] &  (2)  & $-$  & $-$  & $-$   & $-$  \\
2021-11-07\,(59525) & 5.5$\to$3.0  & 139$\pm$33$\to$169\,[$\alpha=-0.33$] &  (2)  &  $-$  & $-$  & $-$   & $-$ \\
2021-10-01\,(59488) & 10.0$\to$3.0  & 115$\pm$24$\to$171\,[$\alpha=-0.33$] &  (2)   & $-$  & $-$  & $-$   & $-$ \\
2021-11-07\,(59526) & 10.0$\to$3.0  & 81$\pm$18$\to$120\,[$\alpha=-0.33$] &  (2)   &  $-$  & $-$  & $-$   & $-$ \\
2023-02-26\,(60002) & 1.70$\to$3.0  & 201$\pm$34$\to$120\,[$\alpha=-0.41$] &  (6)   &  $-$  & $-$  & $-$   & $-$ \\

\hline
\hline
\end{tabular}
\label{tab:obs_results_previous}
\begin{tablenotes}
\item Note: Column (1) lists the observation time in units of YY-MM-DD-HH and MJD. The observed center frequency is present in Column 2. Column 3 gives the flux density with spectral index $\alpha$ if necessary. The spectral index $\alpha$ is used to convert the flux densities at other wavelengths to the frequency used in this study. 
The $\alpha$ of PRS linked to FRB~20121102A is taken from the reference (2). 
The values of $\alpha$ of FRB~20190520B are from the corresponding reference, and the symbol $\ast$ indicates that the $\alpha$ is taken from the mean value of the reference (5). The Column 4 gives the references: (1) \citet{Niu2022Natur606873N}, (2) \citet{Zhang2023ApJ95989Z}, (3) \citet{Marcote2017ApJ834L8M}, (4) \citet{Plavin2022MNRAS5116033P}, (5) \citet{Rhodes2023MNRAS5253626R}, (6) \citet{Bhandari2023ApJ958L19B}.
\end{tablenotes}
\end{table*} 

\subsection{Light Curve of the PRS}

Combining the results of this work (Table\,\ref{tab:obs_results_thiswork}) with previous studies (Table\,\ref{tab:obs_results_previous}), we obtained a total number of 27 measurements for the PRS of FRB~20121102A at L-band 1.5\,GHz and 35 detections for the PRS of FRB~20190520B at S-band 3\,GHz, with observational timescales ranging from 2016 to 2023. 
These observations allow us to investigate the radio light curve and study the flux variability of the persistent radio sources of the two repeat and famous FRBs 20121102A and 20190520B, spanning the longest timescale to date for each target. 
Fig.\,\ref{fig:light_curve_prs121102_t} and \ref{fig:light_curve_prs190520_t} show the radio light curve of the PRSs of FRB\,20121102A and FRB\,20190520B, respectively. 

For the PRS of FRB~121102A at L-band, the mean flux density of the 27 observations is $S_{\rm mean}=235.1\pm34.0\,\mu Jy$, with minimum and maximum values of $S_{\rm min}=110.3\pm22.7\,\mu Jy$ and $S_{\rm max}=361.6\pm56.4\,\mu Jy$, respectively. 
In the 1.5\,GHz observations of 2016, 2017, 2019, and 2023, the average flux densities are $193.6\pm23.2\,\mu Jy$, $247.0\pm33.2\,\mu Jy$, $254.8.2\pm27.0\,\mu Jy$, $251.9\pm45.5\,\mu Jy$, respectively, showing a yearly variation in mean intensity. 
Also, the minimum and maximum flux densities are $168.0\pm11.0\,\mu Jy$ and $220.0\pm40.0\,\mu Jy$ in 2016, $230.0\pm12.0\,\mu Jy$ and $278.0\pm62.0\,\mu Jy$ in 2017, $236.0\pm26.0\,\mu Jy$ and $282.0\pm28.0\,\mu Jy$ in 2019, as well as $110.3\pm22.7\,\mu Jy$ and $361.6\pm59.5\,\mu Jy$ in 2023, indicating yearly variations. 

For the PRS of FRB~20190520B at S-band, the mean flux density of 35 observations is $S_{\rm mean}=179.3\pm22.8\,\mu Jy$, with minimum and maximum values of $S_{\rm min}=111.0\pm33.0\,\mu Jy$ and $S_{\rm max}=233.3\pm18.8\,\mu Jy$, respectively. 
For the 3\,GHz observations in 2020, 2021, and 2023, the average flux densities are $194.1\pm22.9\,\mu Jy$, $147.5\pm37.6\,\mu Jy$, and $183.2\pm12.8\,\mu Jy$, respectively, showing a yearly variation in mean flux density. 
Also, the minimum and maximum flux densities are $160.0\pm21.0\,\mu Jy$ and $233.0\pm29.0\,\mu Jy$ in 2020, $111.0\pm33.0\,\mu Jy$ and $190.0\pm70.0\,\mu Jy$ in 2021, as well as $139.9\pm11.9\,\mu Jy$ and $233.3\pm18.8\,\mu Jy$ in 2023, also showing yearly variations.

Based on observations in Tables\, \ref{tab:obs_results_thiswork} and previous results in Table \ref{tab:obs_results_previous}, we illustrate the flux differences ($\rm \Delta Flux$) as a function of time differences for two epochs of PRS of FRB~20121102A and FRB~20190520B in Fig.\ref{fig:diff_flux_vs_diff_time}. The flat solid blue line shows the linear fit for all points with $1\sigma$ shadows, and the red diamonds show the median values of $\rm \Delta Flux$ derived from every 20 points in the figure. 
The consistent variation observed across different time intervals indicates the presence of variations on both short-term and long-term timescales, consistent with the monthly and yearly variations mentioned above.

A flat structure function indicates that the variability in the luminosity of the PRS shows no significant difference in amplitude over both short and long timescales. In other words, the variability does not exhibit a clear time dependence or characteristic timescale. 
This suggests that within the studied time range, there have been no significant changes in the internal or external environment of the radio source. Such variability may arise from more random physical processes or the actual changes in luminosity may be obscured by noise. Considering the uneven sampling of observation epochs and the limitation of sampling size, the current data cannot rule out other possibilities and draw a conclusion. 

\subsection{Analysis of Variability}

In this section, we quantify the degree and significance of the flux density variability of the two PRSs using the same strategies in the literature and compare the variability with other extragalactic variable sources.

\subsubsection{The $\chi^{2}$ test and the modulation parameter}

Following the same strategy used in \citet{Kesteven1976AJ81919K}, we measure the significance of variability in the radio light curve by calculating the $\chi^{2}$ probability that the intensity of the source remained constant over the observing period. 
Given a set of flux densities $S_{i}$ with error $\sigma^{2}_{i}$ for one PSR as listed in Table\,\ref{tab:obs_results_thiswork} and Table\,\ref{tab:obs_results_previous}, the $\chi_{\rm lc}^{2}$ of the residuals of a weighted fit to a line of constant flux density is computed by the following expressions, 

\begin{equation}
\chi_{\rm lc}^{2} = \sum_{i=1}^{n} \frac{(S_{i} - \tilde{S})^{2}}{\sigma_{i}^{2}}.
\label{chisquared}
\end{equation}

Where $\tilde{S}$ is the weighted mean flux density over the number of $n$ flux density measurements in the radio light curve,
\begin{equation}
\tilde{S} = \sum_{i=1}^{n} \left( \frac{ S_{i}}{\sigma_{i}^{2}} \right) / \sum_{i=1}^{n} \left( \frac{1}{\sigma_{i}^{2}} \right).
\label{w_mean}
\end{equation}

If the errors of the flux densities are drawn from normal random distributions, we expect $\chi_{\rm lc}^{2}$ to be followed by the theoretical distribution $\chi^{2}$ with n-1 degrees of freedom. Then, for the light curve of each source, we can calculate the probability $p(\chi_{\rm lc}^{2})$ of exceeding $\chi_{\rm lc}^{2}$ by chance for a random distribution. 

The value of $p(\chi_{\rm lc}^{2})$ indicates the likelihood that the source has a consistent flux density over the observation time. Thus, a high value of $p(\chi_{\rm lc}^{2})$ suggests a source has a stable flux density over the observation time, rather than showing significant variability. 
Previous studies consider a source to be variable if the probability $p(\chi_{\rm lc}^{2})<0.001$ \citep[e.g., ][]{Kesteven1976AJ81919K,Gaensler2000PASA1772G,Bannister2011MNRAS412634B,Bell2014MNRAS438352B}. 
Hence, the two PRSs in this work are considered to be variables, as their radio light curves exhibit $p(\chi^{2})<0.001$. 
The observed $\chi^{2}$ and $p(\chi^{2})$ of the two PRSs are listed in columns 2 and 3 of Table\,\ref{tab:var_summary}.

The above $\chi^{2}$ analysis is used to test the significance of the variability, and then we use the modulation index to measure the degree of the variability, as below, 
\begin{equation}
m = \frac{\sigma_{S}}{\overline{S}} ,
\label{mod_index}
\end{equation}
\noindent where $\sigma_{S}$ is the standard deviation of the flux density of the source light-curve; and $\overline{S}$ is the mean flux density of the source light-curve. 
The values of $m$ are 0.217$\pm$0.018 and 0.164$\pm$0.017 for the two PRSs of FRB~20121102A and FRB~20190520B, respectively. The error of the modulation index $m$ is determined by performing 1000 repeat randomly generations of the flux values by assuming Gaussian distribution with flux error as standard deviation of the distribution, through equation\,\ref{mod_index}, and please see details in the notes of Table\,\ref{tab:var_summary}.
The degree of variability, as measured by the above modulation index $m$, could include the variation due to the statistical uncertainty in the measurement \citep{Bell2014MNRAS438352B}.
For instance, the light curve for any source with measurement errors on the flux densities will show some variation, even if the source itself is non-varying \citep{Barvainis2005ApJ618108B}. 
To discuss the impact of flux density errors on the degree of the variability $m$, we show the modulation indices as a function of mean signal-to-noise ratios ($\overline{\rm SNR}$) in Fig.\,\ref{fig:snr_vs_m}, reproduced based on results of \citet{Bell2014MNRAS438352B}. 
The dashed line shows the expected random variability due to the measurement uncertainties. 
The lime and yellow stars show the two PRSs in this work and the red stars denote the four variables identified by \citet{Bell2014MNRAS438352B} who did a blind search of variables in radio surveys. 
From Fig.\,\ref{fig:snr_vs_m}, we can see that two PRSs are real variables rather than random variations (dashed line in Fig.\,\ref{fig:snr_vs_m}) derived from the measurement uncertainties of intensity.

To compare with other variables, we also calculated the de-biased modulation index $m_{d}$, which takes into account the errors in the flux measurements (see \citealt{Akritas1996ApJ470706A}; \citealt{Bell2014MNRAS438352B}; \citealt{Sadler2006MNRAS371898S}) and is defined as 

\begin{equation}
 m_{d} = \frac{1}{\overline{S}} \sqrt{\frac{\sum_{i=1}^{n}(S_{i} - \overline{S})^{2} - \sum_{i=1}^{n} \sigma_{i}^{2} }{n}} 
\label{debiased_mod_index}
\end{equation}

For the two PRSs of FRB~20121102A and FRB~20190520B, we obtained $m_{d}$ of 21.5$\pm$1.8\% and 16.0$\pm$1.8\%, together with their redshifts of $z$=0.1927 \citep{Tendulkar2017ApJ834L7T}, and $z$=0.241 \citep{Tendulkar2017ApJ834L7T}, which suggests that they are variable sources by comparing with the relationship of $m_{d}$ and redshift in figure 6 of \citet{Barvainis2005ApJ618108B} for the radio variability of radio-quiet and radio-loud quasars. The error of the de-biased modulation index $m_{d}$ is estimated by conducting 1000 repeated random generations of flux values, as the same strategy used in equation\,\ref{mod_index}. 
\subsubsection{The flux coefficient of variation}

Following the same strategy used in the transient surveys \citep{Sarbadhicary2021ApJ92331S, Andersson2023MNRAS5232219A} and in the radio variability of the PRS associated with FRB \citep{Zhang2023ApJ95989Z}, we estimate the flux coefficient of variation (V) to characterize the long-term variability, based on a series of flux density measurements between 2016 and 2023, 

\begin{equation} \label{eq:V}
    V = \frac{\sigma_{F}}{\overline{F}} = \frac{1}{\overline{F}} \sqrt{\frac{N}{N-1} \left(\overline{F^2} - \overline{F}^2\right)}
\end{equation}

and the significance of the variability can be qualified by $\eta$, which is similar to the weighted reduced $\chi^{2}$ statistic,

\begin{equation} \label{eq:eta}
    \eta = \frac{1}{N-1} \sum_{i=1}^{N} \frac{\left(F_{i} - \xi_{F}\right)^2}{\sigma_{F_{i}}^2}.
\end{equation}
Where $\sigma_{F}$ refers to the standard deviation of the flux density measurements $F_i$, and $\overline{F}$ is the mean value of $F_{i}$. 
$\xi_F$ is the weighted mean of the flux density measurements, defined as $\xi_F = (\sum_{i=1}^{N} F_i/\sigma_{F_{i}}^2)/(\sum_{i=1}^{N} 1/\sigma_{F_{i}}^2)$. 

The coefficients V are 0.221$\pm$0.018 and 0.164$\pm$0.018 for the PRSs of FRB\,20121102A and FRB\,20190520B, with a significance level $\eta$ of 82 and 207, respectively, as listed in Table\,\ref{tab:var_summary}. 
The variability amplitude  $V$=22.1\% of PRS with FRB~121102A is slightly higher than the $V$$\sim$15\% reported in \citet{Rhodes2023MNRAS5253626R}, and the significance of the variability in this study, $\eta=82$, is significantly higher than the $\eta\sim 8$ in \citet{Rhodes2023MNRAS5253626R}. 

Considering the fewer observation epochs and shorter timescales in \citet{Rhodes2023MNRAS5253626R} compared to this study, the amplitude V and significance $\eta$ of the variation in the PRS of FRB~20121102A reported here are more statistically reliable. 
In this study, the V and $\eta$ values of the two PRSs support that they are moderate-variability sources when compared with the V–$\eta$ distribution of 370 variable candidates \citep[see figure 6 of ][]{Sarbadhicary2021ApJ92331S}.  

\subsection{Variability measurements of parsec variable sources}

As discussed in previous studies, the variable components of the PRSs associated with FRB~20121102A \citep{Plavin2022MNRAS5116033P} and FRB~20190520B \citep{Zhang2023ApJ95989Z} have been constrained on parsec scales. 
To compare the flux variability of PRS with other parsec extragalactic variable objects, we adopted the variability parameter $V_{\rm flux}$ that is commonly used in measuring the variability of parsec Active Galactic Nuclei \citep[AGN, ][]{Aller1999ApJ512601A, Jorstad2007AJ134799J, Hodge2018ApJ862151H}, by considering the uncertainties of measurements, 
\begin{equation}
    V_{\rm flux}=\frac{(S_{\rm max}-\sigma_{\rm max})-(S_{\rm min}+\sigma_{\rm min})}{( S_{\rm max}-\sigma_{\rm max})+(S_{\rm min}+\sigma_{\rm min})} 
\end{equation}
Where $S_{\rm max}$ and $S_{\rm min}$ refer to the maximum and minimum values of the measured flux densities with uncertainties of $\sigma_{\rm max}$ and $\sigma_{\rm min}$. 
The parameter $V_{\rm flux}$ serves as a measure of the amplitude variability by taking into account the effect of measurement uncertainty. 
Therefore the variability $V_{\rm flux}$ of PRSs associated with FRB~20121102A is 0.39$\pm$0.05 at L-band, which is close to the peak of $V_{\rm flux}$ distribution of the subclass Flat Spectrum Radio Quasars (FSRQs, $\alpha>-0.5$) in \citet{Hodge2018ApJ862151H}. 
Likewise, the $V_{\rm flux}$=0.21$\pm$0.05 for the PRS associated with FRB~20190520B at S-band, which is also close to the peak of $V_{\rm flux}$ distribution of the subclass FRRQs in \citet{Hodge2018ApJ862151H}. 
This indicates that the host galaxies of the two persistent radio sources associated with the two repeating fast radio bursts exhibit comparable variability to the category of flat-spectrum radio quasars. 

\begin{figure*}[!hpt]
 \centering
   \includegraphics[width = 0.6\textwidth]{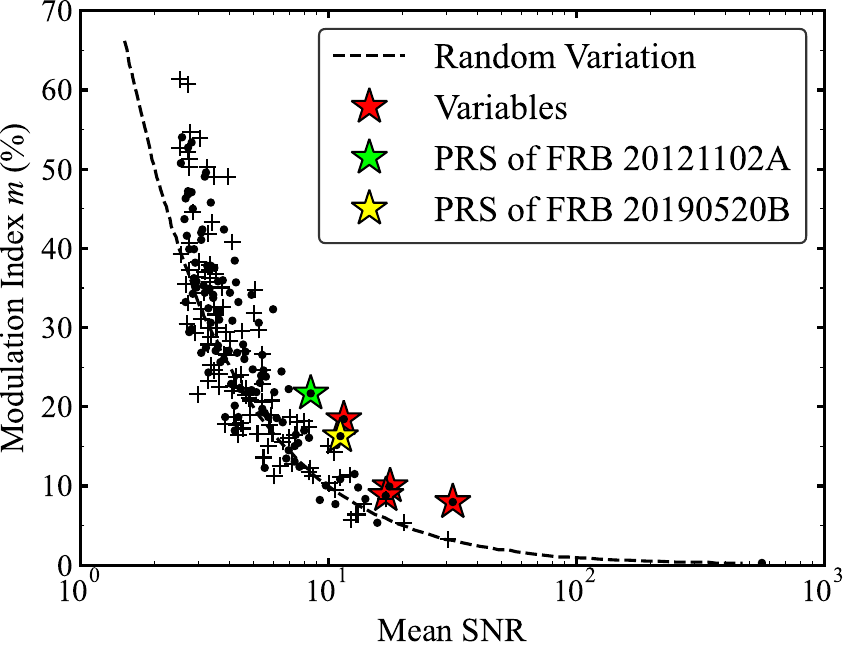}
    \caption{The modulation indices ($m$) as a function of mean single-to-noise ratio ($\overline{\rm SNR}$), reproduced based on results of \citet{Bell2014MNRAS438352B} with permission.
    The dashed line shows the variability that
would be expected from random variations as derived from the measurement
uncertainty. The dots and crosses show the sources observed in different epochs in \citet{Bell2014MNRAS438352B}. 
The red stars denote the four variable sources identified in \citet{Bell2014MNRAS438352B} from a blind search of variables in radio surveys. 
The lime and yellow stars display the positions of the PRSs of FRB~20121102A and FRB~20190520B observed in this work, suggesting that they are variables as defined by \citet{Bell2014MNRAS438352B}. 
Thus, variations in the two PRSs are real rather than random variations derived from measurement uncertainties.}
 \label{fig:snr_vs_m}%
\end{figure*}

\section{Discussion}
\label{sect:discussion}

\subsection{Is the variability intrinsic? }
\label{sect:variable_intrinsic}

Flux variability in radio sources can be attributed to extrinsic effects, such as scintillation of the wavefronts propagate through an inhomogeneous ionized medium \citep[e.g., ][]{Rickett1990ARAA28561R}. 
The two main types of scintillation expected for point sources are: diffractive and refractive. 
Only the latter will be captured in broad-band observations. 
In the following paragraph, we will discuss the refractive scintillation properties of the two PRSs using the models in \citet{Walker1998MNRAS294307W}, and compare it with the observed variability parameters in Sect.\,\ref{sect:results}. 

According to the scintillation of bursts in FRB~20121102A \citep{Michilli2018Natur553182M,Hessels2019ApJ876L23H} and FRB~2019520B \citep{Ocker2022ApJ93187O}, the scintillation of their PRS could be primarily attributed to being the Milky Way.
Based on the distribution of ISM electron density of Milky way from \cite{Cordes2002astroph7156C}, the flux fluctuations from scintillation can be measured by the parameter ``scattering length'' as $\zeta = (\nu_0/\nu)^{17/10}$ \citep{Walker1998MNRAS294307W}, where $\nu_0$ is the transitional frequency between strong scattering regime ($\nu<\nu_0$; $\zeta\gg 1$) and weak scattering regime ($\nu>\nu_0$; $\zeta\ll 1$). 
In the direction of the FRB~20190520B, the transitional frequency $\nu_0 = 12.5$\,GHz as measured by \citet[]{Zhang2023ApJ95989Z}. 
Our observing frequency of 3\,GHz for the associated PRS is thus under the strong scattering regime. The modulation index for refractive scintillation of the point source variations is $m_{p} = \zeta^{-1/3} = (\nu/\nu_0)^{17/30}$ and the timescale of variation $t_r \sim (2\mathrm{hr})(\nu_0/\nu)^{11/5}$ \citep{Walker1998MNRAS294307W}. 
The observed modulation index values $\sim 0.16$ (see $m, m_d, and\, V$ in Table\,\ref{tab:var_summary}) do not reach the expected value of $m_{p}=0.44$ due to scintillation. 
Also, the short variation timescale $t_r \sim 1.9$ days from scintillation cannot explain the long-term variation seen in the radio light curve. 
This is consistent with the result 
in \citet{Zhang2023ApJ95989Z} for the PRS of FRB~20190520B, which also found the 3\,GHz flux variation is not matched with the predicted results from scintillation.
 
%
Likewise, in the direction of the FRB~20121102A, $\nu_0 = 38$\,GHz \citep{Chen2023ApJ958185C}, and our observation taking at 1.5\,GHz for the associated PRS is also under the strong scattering regime. 
The modulation index and the timescale of scintillation are $m_{p} =0.16$ and $\rm t_r\sim 102\,days$. The $m_{p}$ is smaller than the observed modulation index of $\sim$0.22$\pm$0.02 in the radio light curve (see $m, m_d, and\, V$ in Table\,\ref{tab:var_summary}) and the timescale due to scintillation is insufficient to account for the observed monthly--yearly variation in the light curve. 
This is consistent with \citet{Chen2023ApJ958185C} who studied the variability of the PRS of FRB~121102A for $\sim$4 times of VLA observations and suggested that the variation is unlikely to be due to scintillation. 
Therefore, we conclude that the detected variability of the two PRSs observed in this work is more likely to be intrinsic factors rather than extrinsic effects such as scintillation.
However, considering the uncertainties from assumptions and models, we cannot completely rule out the possibility of the variability is due to refractive scintillation, by comparing the timescales and modulation index between models and observations.

\subsection{Obscured Star formation in the host galaxy?}
\label{sect:star_formation_rate}
The host galaxies of the two FRBs observed in this work are suggested to be star-forming dwarf galaxies \citep[e.g.,][]{Chatterjee2017Natur54158C,Niu2022Natur606873N}. 
To investigate the properties of the star-forming host galaxies, we measured the star formation rate (SFR) based on the radio observations and results in this study. 
According to the relation of the SFR and IR emission (Eq. 4 of \citealt{Murphy2011apj737}), together with the tight empirical FIR–radio correlation \citep{Helou1985ApJ298L7H,deJong1985AA147L6D}, one can express the radio-derived SFR at the frequency bands dominated by non-thermal emission \citep{Greiner2016AA593}, assuming the same power law function form (F$_{\nu}$ $\propto$ $\nu^{\alpha}$) and a $k$-correction, 
\begin{equation}
\begin{split}
 \left( \frac{\rm SFR_{radio}}{M_{\odot}\,{\rm yr}^{-1}} \right) = 0.059 \times \left( \frac{F_\nu}{\rm \mu Jy} \right) \times \left( \frac{d_{\rm L}}{\rm Gpc} \right)^2 \\ 
\times \left(\frac{\nu}{\rm GHz}\right)^{-\alpha}(1+z)^{-(\alpha +1)}
 \end{split}
\label{eqn:radioSFR}
\end{equation}

\noindent where $F_{\nu}$ is the observed radio flux density at frequency $\nu$, $d_{\rm L}$ is the luminosity distance, $\alpha$ is the spectral slope of the SED, and $z$ is the redshift of the FRB host. 

For FRB\,20121102A, instead of using the observed flux density from our observations, we derive the mean values of flux density $F_{\rm mean}=235.1\pm34\,\mu Jy$ and spectral index $\alpha_{\rm mean}=-0.48$ from all the observations at 1.5\,GHz as listed in Table\,\ref{tab:obs_results_previous} and Table\,\ref{tab:obs_results_thiswork}. Together with the redshift of $z=0.193$ and distance of $d_{\rm L}\rm =0.972\,Gpc$ \citep{Tendulkar2017ApJ834L7T}, this yields $\rm SFR_{1.5\,GHz}=14.6 \pm 2.1$ $\rm M_{\odot}~{\rm yr^{-1}}$. 
This is consistent with the star formation rate directly derived from the radio luminosity at other radio bands within uncertainties, based on the relation between SRF and radio luminosity in the equation (17) of $\rm SFR_{radio}[M_{\odot}\,yr^{-1}]=6.35\times10^{29}\times L_{radio}[erg\,s^{-1}\,Hz^{-1}]$ in \citet{Murphy2011apj737}, such as $\rm SFR_{1.28\,GHz}\sim 16.51\,M_{\odot}/yr$ ( $\rm L_{1.28\,GHz}\sim 2.6\times 10^{29}\,erg\,s^{-1}\,Hz^{-1}$ in \citealt{Rhodes2023MNRAS5253626R}), and $\rm SFR_{1.7\,GHz}\sim 11.2\,M_{\odot}/yr$ ($\rm \rm L_{1.7\,GHz}\sim 1.76\times 10^{29}\,erg\,s^{-1}\,Hz^{-1}$ in \citealt{Marcote2017ApJ834L8M}). 
The radio-derived SFR is a factor of $\sim$44 higher than the 
optical inferred SFR of $\sim$0.4\,$\rm M_{\odot}\,yr^{-1}$ from \citet{Tendulkar2017ApJ834L7T} after extinction correction. 
The 3$\sigma$ ($51\,\mu Jy$ ) non-detection of the PRS of FRB~20121102A at 230\,GHz with ALMA gives an upper limit of SFR $\sim 12.8\,\rm M_{\odot}\,yr^{-1}$, by following the equation (2) in \citet{Carilli1999ApJ513L13C} as outlined in \citet{Chatterjee2017Natur54158C}. 
The radio SFR derived from radio centimeter is found to be even higher than the upper limit of SFR derived from the ALMA detection limit at submillimeter, which could be due to (1) the large uncertainties in the measurement of SFR; (2) a larger fraction of radio emission from the non-star-formation process, as previously suggested by \citet{Kennicutt1989AJ971022K}. 
In total, the host galaxy of FRB~20121102A is likely to be highly obscured or the observed radio emission is likely to include a significant portion of non-star-forming emissions. 
The latter is consistent with the European VLBI Network (EVN) observations with angular resolutions of $\rm \sim 2\,mas$ in \citet{Marcote2017ApJ834L8M} and \citet{Plavin2022MNRAS5116033P} who suggest that the radio emission of the PRS is not related to star formation processes.

For FRB\,20190520B, instead of using the observed flux density from our observations, we derive the mean values of flux density $S_{\rm mean}=179.3\pm22.8\,\mu Jy$ and spectral index $\alpha_{\rm mean}=-0.37$ from all the observations at 3.0\,GHz as listed in Table\,\ref{tab:obs_results_previous} and Table\,\ref{tab:obs_results_thiswork}. Together with the redshift of $z=0.241$ and distance of $d_{\rm L}\rm =1.218\,Gpc$ \citep{Niu2022Natur606873N}, this gives $\rm SFR_{radio}=20.6 \pm 2.6$ $\rm M_{\odot}~{\rm yr^{-1}}$. 
This is consistent with the star formation rate directly derived from the radio luminosity at other observations within uncertainties, such as $\rm SFR_{3\,GHz}\sim 19.1\,M_{\odot}/yr$ ( $\rm L_{3\,GHz}\sim 3\times 10^{29}\,erg\,s^{-1}\,Hz^{-1}$ in \citealt{Niu2022Natur606873N}), and $\rm SFR_{1.5\,GHz}\sim 29.8\,M_{\odot}/yr$ ($\rm \rm L_{1.5\,GHz}\sim 4.7\times 10^{29}\,erg\,s^{-1}\,Hz^{-1}$ in \citealt{Marcote2017ApJ834L8M}). 
We noted that the radio-derived SFR is much higher than the star formation rate of $\sim$0.4\,$\rm M_{\odot}~{\rm yr^{-1}}$ from the optical $H_{\alpha}$ luminosity in \citet{Niu2022Natur606873N}, with a ratio of $\rm SFR_{\rm optical}/SFR_{\rm radio}\sim 51$. This suggests that the host galaxy of FRB\,20190520B is largely obscured by dust. 
Compared to the SFR ratio between radio and optical for the other four FRBs with hosts (see figure 6 of \citealt{Dong2023arXiv230706995D}), the FRB host of FRB~20190520B is likely to be the most dust-obscured star formation in the host galaxies of FRBs. 
However, we cannot exclude that most radio emissions are not from the star formation process as the observed flux density from the EVN observations with angular resolutions of $\rm \sim 2\,mas$ in \citet{Bhandari2023ApJ958L19B} are comparable with the intensity from VLA observations.

\subsection{Constrain the magnetic fields }

The observed timescale of the PRS flux evolution might be used to constrain the scale of the PRS region. As shown in Figs.\,\ref{fig:light_curve_prs121102_t}, \ref{fig:light_curve_prs190520_t}, and \ref{fig:snr_vs_m}, and the typical timescale of the PRS flux evolution is about $\Delta t\sim{\rm a~few~days}$, which constrains the PRS emission region to $R\sim c\Delta t\sim10^{-3}~{\rm pc}(\Delta t/1~{\rm days})$. Such a small scale might lead to the free-free absorption by the non-relativistic electrons from the PRS region \citep{Yang2020ApJ8957Y}. 
The free-free optical depth is given by
\begin{align}
\tau_{\rm ff}=\alpha_{\rm ff}R=0.018T^{-3/2}Z^2n_en_i\nu^{-2}\bar{g_{\rm ff}}R,
\end{align}
where $n_e$ and $n_i$ are the number densities of electrons and ions, respectively (Here we assume that $n_e=n_i$ and $Z=1$ for the emission region), and $\bar{g_{\rm ff}}\sim1$ is Gaunt factor, $T$ is the thermal gas temperature. Since FRBs with GHz frequency are observable, the transparency condition requires $\tau_{\rm ff}<1$, leading to 
\begin{align}
n_e<1.3\times10^5~{\rm cm^{-3}}\left(\frac{T}{10^4~{\rm K}}\right)^{3/4}\left(\frac{\nu}{1~{\rm GHz}}\right)\left(\frac{R}{10^{-3}~{\rm pc}}\right)^{-1/2}.\nonumber\\
\label{eq:ff}
\end{align}
The DM contributed by the PRS emission region satisfies ${\rm DM}\sim n_e R\lesssim130~{\rm pc~cm^{-3}}$. 
The parallel magnetic field strength could be constrained by the observed RM, 
\begin{align}
B_\parallel\simeq1.23~{\rm \mu G}\frac{{\rm RM}}{n_eR}\frac{{\rm pc~cm^{-3}}}{{\rm rad~m^{-2}}}\gtrsim95~{\rm \mu G}\left(\frac{{\rm RM}}{10^4~{\rm rad~m^{-2}}}\right)
\end{align}
for $\nu\sim1~{\rm GHz}$ and $T\sim10^4~{\rm K}$. For FRB~121102 with ${\rm RM}\sim 10^5~{\rm rad~m^{-2}}$, the constrained magnetic field is $B_\parallel\gtrsim1~{\rm mG}$. For FRB~190520 with ${\rm RM}\sim 10^4~{\rm rad~m^{-2}}$, the constrained magnetic field is $B_\parallel\gtrsim0.1~{\rm mG}$.

\section{Conclusion} 
\label{sect:conclusion}
In this study, we present 22 observations of two persistent radio sources associated with two repeats of fast radio bursts FRB~20121102A and FRB~20190520B using VLA in 2023. 
We measured the radio continuum properties of the two persistent radio sources. 
Together with the previous measurements, we constructed the radio light curves for PRSs of FRB~20121102A with 27 measurements and FRB~20190520B with 35 observations, spanning the longest timescale recorded to date. 
We adopted several commonly used methods to analyze and discuss the significance and degree of the flux density variability for the two persistent radio sources for observations between 2016 and 2023. The results from all methods suggest that the intensity variabilities of two persistent radio sources are significant and intrinsic. 
The observed variability of the two PRSs exhibits no significant difference in amplitude across both short- and long-term time intervals, indicating that such variability may arise from more random physical processes. 
The degree of variability of the two PRSs is comparable to other parsec radio variable sources such as Flat Spectrum Radio Quasars. We found the radio-derived star formation rates of the host galaxies of the two FRBs (e.g., FRB~20190520B and FRB~20121102A) are one order of magnitude higher than SFR measured by optical $H_{\alpha}$ lines, suggesting that the hosts are highly obscured by dust or a significant fraction of radio emission is from the non-star-forming process. The observed timescale of PRS flux evolution helps constrain the magnetic field of FRB~20121102A with $B_\parallel\gtrsim1~{\rm mG}$ and FRB~20190520B with $B_\parallel\gtrsim0.1~{\rm mG}$. 

\begin{acknowledgments}
We would like to thank the referee for the helpful comments and suggestions on our manuscript. 
AYY acknowledges support from the National Key R$\&$D Program of China No. 2023YFC2206403 and the National Natural Science Foundation of China (NSFC) grants No. 11988101 and No. 12303031.  
Y.F. is supported by National Natural Science Foundation of China grant No. 12203045, by the Leading Innovation and Entrepreneurship Team of Zhejiang Province of China grant No. 2023R01008, and by Key R\&D Program of Zhejiang grant No. 2024SSYS0012. 
DL is a New Cornerstone investigator. 
P.W. acknowledges support from the NSFC Programs No.11988101, 12041303, the CAS Youth Interdisciplinary Team, the Youth Innovation Promotion Association CAS (id. 2021055), and the Cultivation Project for FAST Scientific Payoff and Research Achievement of CAMS-CAS.
CWT and DL acknowledge support from International Partnership Program of Chinese Academy of Sciences, Program No.114A11KYSB20210010.
Ju-Mei Yao was supported by the National Science Foundation of Xinjiang Uygur Autonomous Region (2022D01D85), the Major Science and Technology Program of Xinjiang Uygur Autonomous Region (2022A03013-2), the Tianchi Talent project, and the CAS Project for Young Scientists in Basic Research (YSBR-063), and the Tianshan talents program (2023TSYCTD0013).
The National Radio Astronomy Observatory is a facility of the National Science Foundation, operated under a cooperative agreement by Associated Universities, Inc.
This document was prepared using the collaborative tool Overleaf available at: 
\url{https://www.overleaf.com/}.
 \software{astropy \citep{Astropy2013AA558A33A}, AEGEAN \citep{Hancock2012MNRAS4221812H}, OBIT \citep{Cotton2008PASP439C} }
\end{acknowledgments}

\bibliographystyle{aasjournal}
\bibliography{frb_ref}{}

\end{document}